\documentclass{article}
\usepackage[utf8]{inputenc}
\usepackage[left=3cm, right=3cm, top=4cm, bottom=4cm]{geometry}

\usepackage{amsmath}
\usepackage{amsfonts}
\usepackage{amssymb}

\usepackage[backend=biber, style=ieee]{biblatex}
\bibliography{literature}

\usepackage{bm}
\usepackage{graphicx}

\title{Orthogonal Matching Pursuit \\ with Tikhonov and Landweber Regularization}
\author{Robert Seidel\thanks{R. Seidel is with the Institut für Mathematik, Technische Universität Berlin. He was supported by the German Academic Exchange Service and the Taiwanese Ministry of Science and Technology through the \textit{Taiwan-Germany Summer Institute Program}. The author wishes to thank his supervisor An-Yeu (Andy) Wu, and all reviewers for their comments on this publication.}}
\date{}

\DeclareMathOperator*{\argmax}{arg~max}
\DeclareMathOperator*{\argmin}{arg~min}
\newcommand{\E}{\mathbb{E}}
\newcommand{\N}{\mathbb{N}}
\newcommand{\R}{\mathbb{R}}

\DeclareMathOperator{\rank}{rank}
\DeclareMathOperator{\supp}{supp}
\DeclareMathOperator{\im}{Im}
\newcommand{\op}{\text{op}}

\usepackage{algorithm}

\usepackage{amsthm}
\newtheorem{counter}{counter}
\newtheorem{definition}[counter]{Definition}
\newtheorem{theorem}[counter]{Theorem}

\newlength{\figwidth}
\begin{document}
\maketitle

\begin{abstract}
	The Orthogonal Matching Pursuit (OMP) for compressed sensing iterates over a scheme of support augmentation and signal estimation. We present two novel matching pursuit algorithms with intrinsic regularization of the signal estimation step that do not rely on \textit{a priori} knowledge of the signal's sparsity. An iterative approach allows for a hardware efficient implementation of our algorithm, and enables real-world applications of compressed sensing. We provide a series of numerical examples that demonstrate a good performance, especially when the number of measurements is relatively small.
\end{abstract}


\section{Introduction}

A well-known theorem in signal processing is the Nyquist-Shannon theorem: It states that any band-limited signal can be exactly recovered by sampling with a rate of no more than two times its highest frequency. In many applications, however, signals only carry little information compared to the space where the signal is acquired in. Compressed sensing---we refer to \cite{CW08} for a review---relies on the assumption that the gap between the dimension of the signal and its information content is expressed through the signal's sparsity, i.e. the number of its non-zero entries.
Compressed sensing has demonstrated its capabilities in many applications such as medical imaging \cite{LDP07},
wireless communications \cite{LT10}, and the Internet of Things \cite{LXW13}, see also \cite{QBIN13} for a review.

\subsection{Compressed sensing and the restricted isometry property}
The sampling process itself is described by the application of a linear functional to the signal. In the discrete setting, the acquisition of $m$ samples of an unknown signal $x \in \R^N$ can be written as
$$Ax = y$$
with a \emph{sampling matrix} $A \in \R^{m \times N}$. We say that $x \in \R^N$ is $k$-sparse if $\|x\|_0 := |\supp(x)| = k$. Given a set $S \subset \{1,...,N\}$ with $|S| = k$, we denote by $A_S \in \R^{m \times k}$ the matrix that consists of the columns of $A$ indexed by $S$. Similarly, we denote by $x_S \in \R^k$ the vector that consists of the elements of $x$ indexed by $S$.

It is easy to see that every $k$-sparse signal $x \in \R^N$ cannot be reconstructed by $m < 2k$ measurements. Indeed, given a sampling matrix $A \in \R^{m \times N}$ with $m < 2k$, a basic result of linear algebra yields
$$\rank(A) \leq m < 2k,$$
and one can find a $2k$-sparse vector $w \in \R^N$ with $Aw = 0$. Decomposing $w$ into two $k$-sparse vectors $x,y \in \R^N$ with $w=x-y$, yields
$$Ax - Ay = A(x-y) = Aw = 0,$$
i.e. $Ax = Ay$. We have shown that $A$ does not map all pairs of different $k$-sparse signals to pairs of different samples, and exact signal recovery is impossible.

The previous example shows that $A$ should be injective, at least on the set of sparse signals. This idea motivates the following definition due to \citeauthor{CT05} \cite{CT05}.
\begin{definition}
	Let $A \in \R^{m \times N}$ and $1 \leq k \leq N$. Then, the \emph{restricted isometry constant (RIC)} $\delta_k$ of $A$ is defined as the smallest $\delta \geq 0$ such that
	\begin{equation} \label{eqn:ric}
		(1-\delta_k) \|x\|_2^2 \leq \|Ax\|_2^2 \leq (1+\delta_k) \|x\|_2^2
	\end{equation}
	for all $k$-sparse $x \in \R^N$.
\end{definition}

The restricted isometry constant of order $k$ is hard to verify for a given sampling matrix $A \in \R^{m \times N}$ as it requires the computation of $N! / (N-k)!$ submatrices consisting of $k$ columns taken from $A$. Surprisingly, many random matrices have a small RIC with a high probability. If, for example, the entries of $A$ are sampled independently from a standard Gaussian distribution, the matrix $\sqrt{m} A$ has a small RIC $\delta_k$ of order $k$ with high probability, if
\begin{equation} \label{eqn:ric-m}
	m \geq C k \log(N/k).
\end{equation}
We refer to \cite{CT06} for more details.

\subsection{Signal reconstruction without noise}

Naturally, conditions on the sensing matrix and reconstruction algorithms come in pairs. One famous example due to \citeauthor{CT05} is the basis pursuit (BP) \cite{CT05, C08}: If $A \in \R^{m \times N}$, $1 \leq k \leq N$, and $\delta_{2k} < \sqrt{2}-1$, then every $k$-sparse $x \in \R^N$ is the unique solution of
$$\min_{z \in R^N} \|z\|_1 \text{ s.t. } Az=y$$
with input $y=Ax$. The bound on $\delta_{2k}$ has been further improved in \cite{F10}. The BP can be understood as the convex relaxation of an $\ell_0$-pseudonorm functional \cite{T06}.

Another class of reconstruction algorithms are greedy pursuit algorithms that can be motivated by the following: If the sensing matrix $A$ was an isometry, then $A^{-1} = A^*$, and the signal $x$ can be recovered from $A^*y = A^*Ax = x$. In this ideal scenario, the support of $x$ can be recovered by $\supp(x) = \supp(A^*y)$. If $A$ is not an isometry, we call $u = A^*y$ the \emph{observation vector} of $y$. The orthogonal mating pursuit (OMP) due to \citeauthor{TG07} \cite{TG07} iteratively adds the coordinate of the biggest value in magnitude of $u$ to the recovered support set $S \subset \{1,...,N\}$ of $x$. The new signal estimate $x'$ is then computed by projecting $y$ onto the column space of $A_S$, and the next observation is given by $A^*r$, where $r = y - Ax'$ is the residual. The OMP is summarized in Algorithm~\ref{alg:omp}.

\begin{algorithm} 
	\caption{OMP} \label{alg:omp}
	
	Input: sensing matrix $A \in \R^{m \times N}$, samples $y \in \R^m$, sparsity level $k$.

	Initialize: $r \leftarrow y$, $S \leftarrow \{\}$, $x \leftarrow 0 \in \R^N$.
	
	For $k$ iterations:
	\begin{enumerate}
		\item[] \hspace*{-2em}\textsl{Part 1: Support augmentation}
		\item Observe $A^*y$ and find the index $s$ of the largest element in magnitude, i.e.
			$$s \leftarrow \argmax_{j=1,...,N} |\langle A_j, r\rangle|.$$
		\item Add the element to the support, i.e. $S \leftarrow S \cup \{s\}$.
		\item[] \hspace*{-2em}\textsl{Part 2: Project measured signal}
		\item Obtain new signal estimate by
			\begin{align*}
				x_S &\leftarrow \argmin_{z \in \R^{|S|}} \|y - A_S z\|_2, \\
				x_{S^C} &\leftarrow 0.
			\end{align*}

		\item Update the residual by $r \leftarrow y - A_S x_S$.
	\end{enumerate}
\end{algorithm}

Note that Step~3 of OMP is equivalent to computing the pseudoinverse $A_S^\dagger$, provided that $A_S$ has a trivial kernel. In particular, the new signal estimate is obtained by
\begin{align*}
x_S &\leftarrow A_S^\dagger y, \\
x_{S^C} &\leftarrow 0.
\end{align*}

\citeauthor{TG07} proved the following non-uniform recovery result for the OMP algorithm \cite{TG07}: Given a Gaussian sensing matrix $A \in \R^{m \times N}$, the support of every $k$-sparse signal $x \in \R^N$ is recovered by OMP with input $y = Ax$ with high probability, if
\begin{equation} \label{eqn:omp-m}
	m \geq C k \log N
\end{equation}
holds. Using the restricted isometry condition, \citeauthor{WZWT17} \cite{WZWT17} derived a uniform sharp condition for exact support recovery with OMP: If $A \in \R^{m \times N}$ satisfies
\begin{equation} \label{eqn:omp-ric}
	\delta_{k+1} < \frac{1}{\sqrt{k+1}},
\end{equation}
then the support of every $k$-sparse signal $x \in \R^N$ is exactly recovered by OMP with input $y = Ax$ within $k$ iterations. Conversely, for every sparsity level $k$, there exist a $k$-sparse signal $x \in \R^N$, and a sensing matrix $A \in \R^{m \times N}$ with
$$\delta_{k+1} = \frac{1}{\sqrt{k+1}},$$
such that OMP cannot recover the support of $x$ within $k$ iterations.

Compared to solving the minimization problem of BP, matching pursuit algorithms are known to have a low computational complexity \cite{NV09, TG07}: They add exactly one coordinate per iteration to the support estimate and solve the projection problem for the new signal estimate. They are therefore good candidates for efficient hardware implementations.

\subsection{Signal reconstruction in the presence of noise}

A naturally arising question is the robustness of signal recovery in the presence of noise. In this scenario, the signal acquisition reads as
$$y = Ax + v,$$
where $v \in \R^m$ is an unknown noise term. There are several robustness results for BP \cite{CRT06, CDS98, CT06} and OMP \cite{WZWT17, CW11, W15}. In particular, for the OMP algorithm, a second assumptions besides (\ref{eqn:omp-ric}) needs to be made to guarantee successful signal recovery. Namely, if $A \in \R^{m \times N}$ satisfies (\ref{eqn:omp-ric}) and
$$\min_{i \in \supp(x)} |x_i| > \frac{2 \varepsilon}{1 - \sqrt{K+1} \delta_{k+1}},$$
then the support of every $k$-sparse signal $x \in \R^N$ is recovered by OMP with input $y = Ax$ with the stopping rule $\|r\|_2 \leq \varepsilon$ within $k$ iterations, where $r$ is the OMP residual (see Algorithm~\ref{alg:omp}), and $\varepsilon \geq \|v\|_2$ is the noise energy. A similar \textit{sufficient} condition for exact recovery can be found, see \cite{WZWT17} for more details.

Similar to the basis pursuit, regularization has been introduced to the orthogonal matching pursuit by various means. We have identified two major strands in the available literature: The refinement of the support set augmentation and the regularization of the projection step. Most of the literature focuses on regularization of the support augmentation, i.e. Part~1 in Algorithm~\ref{alg:omp}:
\begin{itemize}
	\item The regularized OMP (ROMP) due to \citeauthor{NV09} \cite{NV09, NV10} computes the observation vector $u = A^*r$ of the residual $r$, and selects up to $k$ support indices from a trusted interval of coefficient magnitude. Namely, ROMP will select a set $J \subset \{1,...,N\}$ of the $k$ largest coefficients in magnitude, and seek a subset $J_0 \subseteq J$ such that the smallest coefficient is not bigger than twice the largest coefficient selected, i.e.
	$$|u_i| \leq 2 |u_j| \text{ for all } i,j \in J_0.$$
	If there are multiple such sets $J_0$, ROMP will choose the one with the maximal energy $\|u_{J_0}\|_2$. The set $J_0$ is then added to the set $S$, which completes the support augmentation. The remainder of the algorithm is similar to the OMP.
	
	\item The compressive sampling matching pursuit (CoSaMP) due to \citeauthor{NT09} \cite{NT09} uses the coordinates $T \subset \{1,...,N\}$ of the $2k$ largest coefficients in magnitude of the observation vector $u = A^*r$ as estimate for $\supp(x)$. A $2k$-sparse signal $\xi \in \R^N$ is then estimated by the least squares problem
	$$\xi_T = \argmin_{z \in \R^{2k}} \|A_T z - y\|_2,$$
	and a $k$-sparse approximation for $x$ is given by the $k$ largest entries of $\xi$ in magnitude. The remainder of the algorithm is similar to the OMP.
	
	\item Finally, the hard thresholding pursuit (HTP) due to \citeauthor{F11} \cite{F11} selects the coordinates of the $k$ largest coefficients in $x + A^*(y - A x)$ as support estimate $S$, and then projects $y$ onto $\im(A_S)$. These steps are iterated with an arbitrary $k$-sparse initialization for $x \in \R^N$ until a halting criterion is met.
\end{itemize}

For a fixed sparsity level $k$, the reconstruction error of these algorithms is linear in $\|v\|_2$, see \cite{NV10, NT09, F11}. The same holds true for the BP \cite{CRT06}.

To the author's best knowledge, there is only one algorithm that applies regularization to the signal estimation in OMP (Step~3 in Algorithm~\ref{alg:omp}): The stochastic gradient pursuit (SGP) algorithm due to \citeauthor{LCHW17} \cite{LCHW17} replaces the computation of the pseudoinverse in Step~3 with the least mean squares (LMS) estimate of $A_S^\dagger y$. This approach is motivated by the LMS adaptive filter due to \citeauthor{WMLJ76} \cite{WMLJ76}, whereby the rows $u_1, ..., u_m \in \R^{1 \times |S|}$ of $A_S$ are considered to be the stochastic input to a digital filter with desired outputs $y \in \R$, and $x_S$ plays the role of that filter's unknown weight vector. The objective of the LMS algorithm is to minimize the mean-squared error
$$
	\E\left[(y_\ell - u_\ell x_S)^2\right] = \E\left[y_\ell^2\right] - 2 \E\left[y_\ell u_\ell\right] x_S + x_S^T \E\left[u_\ell^T u_\ell\right] x_S
$$
by stochastic gradient descent, where the expectation is taken over all rows of $A_S$ with equal probability. The unknown expectations in the gradient
$$\nabla_{x_S} \E\left[(y_\ell - u_\ell x_S)^2\right] = - 2 \E\left[y_\ell u_\ell\right] + 2 \E\left[u_\ell^T u_\ell\right] x_S$$
are approximated by gradients of single samples. We refer to \cite{WMLJ76} for more details. The SGP is summarized in Algorithm~\ref{alg:sgp}.

\begin{algorithm}
	\caption{SGP} \label{alg:sgp}
	Input: sensing matrix $A \in \R^{m \times N}$, samples $y \in \R^m$, residual threshold $\tau$, sparsity estimate $k_\text{max}$.
	
	Initialize: $r \leftarrow y$, $S \leftarrow \{\}$, $x \leftarrow 0 \in \R^N$, $\mu = \frac{2 m}{3 k_\text{kmax}}$.
	
	Until $\|r\|_2 \leq \tau$:
	\begin{enumerate}
		\item Observe $A^*y$ and find the index $s$ of the largest element in magnitude, i.e.
		$$s \leftarrow \argmax_{j=1,...,N} |\langle A_j, r\rangle|.$$
		\item Add the element to the support, i.e. $S \leftarrow S \cup \{s\}$.
		\item Obtain new signal estimate by LMS iteration:
		\begin{enumerate}
			\item Initialize $z_0 \leftarrow x_S \in \R^{|S|}$.
			\item For $\ell = 1,...,M$:
				\begin{align*}
				a_\ell &\leftarrow A_S[\ell,:] \in \R^{1 \times |S|} \quad \text{(the $\ell$-th row of $A_S$)} \\
				d_\ell &\leftarrow y_\ell \in \R \\
				e_\ell &\leftarrow d_\ell - a_\ell \cdot z_{\ell-1} \in \R \\
				z_\ell &\leftarrow z_{\ell-1} + \mu \cdot e_\ell \cdot a^T \in \R^{|S|}
				\end{align*}
			\item Then, update $x_S$ by
				\begin{align*}
				x_S &\leftarrow z_M, \\
				x_{S^C} &\leftarrow 0.
				\end{align*}
		\end{enumerate}
		\item Update the residual by $r \leftarrow y - A_S x_S$.
	\end{enumerate}
\end{algorithm}

Note that the regularization of the support augmentation relies on knowledge about the sparsity level $k$ of the unknown signal $x$. This information is required in every iteration when the next support set estimate of appropriate size has to be determined. In contrast, both OMP and SGP add no more than one coordinate per iteration to the support estimate, and
the SGP algorithm uses the sparsity information only to find an upper bound for the LMS step size $\mu$, see Algorithm~\ref{alg:sgp}.


\subsection{Contributions}

The contributions of this work are as follows:
\begin{enumerate}
	\item Introduce two other means of regularization in the signal estimation step: We could identify only one publication \cite{LCHW17} where the signal estimation step of OMP is regularized. We broaden this picture and introduce two other well-established regularization methods in the signal estimation step of the OMP, namely Tikhonov regularization and Landweber iteration. This approach does not rely on \textit{a priori} knowledge of the signal's sparsity level.
	
	\item Hardware feasibility: In many application areas, limitations of sensor size and battery lifetime create the need for efficient hardware implementations \cite{SM12}. The SGP algorithm's LMS estimate is tailored to be a hardware efficient reconstruction method \cite{LCHW17}. This work broadens the class of iterative hardware efficient algorithms where regularization an intrinsically built-in feature of the proposed algorithm.
	
	\item Provide uniform comparison: This work provides a series of numerical experiments that serve as a proof of concept for the previously made claims. We thereby employ the same stopping criterion for all algorithms, making a fair comparison is possible. As a by-product, this stopping criterion improves the stopping criterion for SGP as proposed in \cite{LCHW17}.
\end{enumerate}

\section{Regularization of the pseudoinverse}

Most of the reviewed literature employs regularization in the support augmentation step of OMP, i.e. Part~1 of Algorithm~\ref{alg:omp}. This work aims to regularize the computation of the signal estimate, i.e. Part~2 of Algorithm~\ref{alg:omp}. We assume throughout this whole section that $|S| \leq m$, and $\ker A_S = \{0\}$. The latter holds true with high probability for many matrices with random entries, see e.g. \cite{BG09}.

We begin with the noise-free scenario. Given the exact measurements $y$ and a support set estimate $S$, the signal estimate $x$ is given by
\begin{equation*}
	\begin{aligned}
		x_S &= \argmin_{z \in \R^{|S|}} \|y - A_S z\|_2 = A_S^\dagger y, \\
		x_{S^C} &= 0.
	\end{aligned}
\end{equation*}
If $S = \supp x$, it is easy to see that the signal estimate corresponds to the true signal. If only an approximation $y^\varepsilon$ of $y$ with $\|y^\varepsilon - y\|_2 \leq \varepsilon$ for $\varepsilon > 0$ is available, the OMP signal estimate reads as
\begin{equation} \label{xS}
	\begin{aligned}
		x_S^\varepsilon &= \argmin_{z \in \R^{|S|}} \|y^\varepsilon - A_S z\|_2 = A_S^\dagger y^\varepsilon \\
		x_{S^C}^\varepsilon &= 0.
	\end{aligned}
\end{equation}
It is a standard result from numerical linear algebra that, depending on the condition number $\kappa$ of the matrix $A_S$, the measurement error $y^\varepsilon - y$ amplifies by the action of the pseudoinverse. The condition number is defined as
$$\kappa = \frac{\sigma_\text{max}(A_S)}{\sigma_\text{min}(A_S)},$$
where $\sigma_\text{max}(A_S)$ and $\sigma_\text{min}(A_S)$ are the biggest and smallest non-zero singular values of $A_S$, respectively. If $A$ has the restricted isometry constant $\delta_k$ of order $k \geq |S|$, equation (\ref{eqn:ric}) implies that the singular values of $A_S$ lie between $\sqrt{1-\delta_k}$ and $\sqrt{1+\delta_k}$, and therefore
$$\kappa \leq \frac{\sqrt{1+\delta_k}}{\sqrt{1-\delta_k}}.$$

In the light of (\ref{eqn:omp-m}), a favorable condition number is reached, when the number of measurements is large, compared to the ambient dimension $N$ and the RIC-order $k$. The effect of regularizing the signal estimate in the OMP is therefore stronger in the regime where $m < C k \log (N/k)$.

There are two standard approaches due to Tikhonov and Landweber that deal with unstable solutions under data perturbations. One famous regularization method is Tikhonov regularization, where the estimate $x^\varepsilon$ is replaced by
\begin{equation} \label{xS-tikhonov}
	\begin{aligned}
		x_S^{\alpha, \varepsilon} &= (A_S^*A_S + \alpha I)^{-1} A_S^*y^\varepsilon, \\
		x_{S^C}^{\alpha, \varepsilon} &= 0,
	\end{aligned}
	\end{equation}
for some $\alpha > 0$.

The regularized signal estimate is an approximation of the pseudoinverse in the following sense \cite{EHN00}:
\begin{theorem}
	Let $x^{\alpha, \varepsilon}$ be defined as in (\ref{xS-tikhonov}) with $S = \supp x$. Let $y = Ax$ and $\|y^\varepsilon - y\|_2 \leq \varepsilon$. For $\alpha > 0$, the noise amplification is controlled by the regularization parameter as
	$$\|x^{\alpha,0} - x^{\alpha,\varepsilon}\| \leq \frac{\varepsilon}{\sqrt \alpha}.$$ If $\alpha = \alpha(\varepsilon)$ is such that
	$$\lim_{\varepsilon \to 0} \alpha(\varepsilon) = 0 \quad \text{and} \quad \lim_{\varepsilon \to 0} \frac{\varepsilon^2}{\alpha(\varepsilon)} = 0,$$
	then
	$$\lim_{\varepsilon \to 0} x^{\alpha(\varepsilon), \varepsilon} = A^\dagger y.$$
\end{theorem}


A straightforward extension of the signal estimation of the OMP algorithm is to replace the direct signal estimate (\ref{xS}) by its regularized counterpart (\ref{xS-tikhonov}). The Tikhonov regularized OMP (T-OMP) is summarized in Algorithm~\ref{alg:t-omp}. Note that the T-OMP algorithm relies on the computation of a matrix inverse, which may prevent a hardware efficient implementation.
\begin{algorithm}
	\caption{T-OMP} \label{alg:t-omp}
	Input: sensing matrix $A \in \R^{m \times N}$, samples $y \in \R^m$, noise level $\varepsilon$, regularization parameter $\alpha > 0$.
	
	Initialize: $r \leftarrow y$, $S \leftarrow \{\}$, $x \leftarrow 0 \in \R^N$.
	
	Until $\|r\|_2 \leq \varepsilon$:
	\begin{enumerate}
		\item Observe $A^*y$ and find the index $s$ of the largest element in magnitude, i.e.
		$$s \leftarrow \argmax_{j=1,...,N} |\langle A_j, r\rangle|.$$
		\item Add the element to the support, i.e. $S \leftarrow S \cup \{s\}$.
		\item Obtain new signal estimate by
		\begin{align*}
			x_S &\leftarrow (A_S^*A_S + \alpha I)^{-1} A_S^*y, \\
			x_{S^C} &\leftarrow 0.
		\end{align*}
		
		\item Update the residual by $r \leftarrow y - A_S x_S$.
	\end{enumerate}
\end{algorithm}

Another well-known regularization method is the Landweber iteration. The new signal estimate reads as
\begin{equation} \label{xS-landweber}
	\begin{aligned}
		x_S^{\ell, \varepsilon} &= x_S^{\ell-1, \varepsilon} + \omega A_S^*(y^\varepsilon - A_S x_S^{\ell-1, \varepsilon}), \\
		x_{S^C}^{\ell, \varepsilon} &= 0,
	\end{aligned}
\end{equation}
$$$$
where $0 < \omega \leq \|A_S\|^{-2}_\text{op}$, and $x^{0,\varepsilon} = 0$. The Landweber iteration is an approximation of the pseudoinverse in the following sense \cite{EHN00}:

\begin{theorem}
	If $y = Ax$, $\|y^\varepsilon - y\|_2 \leq \varepsilon$ and $(x^{\ell,0})_\ell$, $(x^{\ell, \varepsilon})_\ell$ are two iteration sequences given by (\ref{xS-landweber}), then
	$$x^{\ell,0} \to A^\dagger y \text{ as } \ell \to \infty, \quad \text{and} \quad \|x^{\ell, 0} - x^{\ell, \varepsilon}\| \leq \sqrt{\ell\,} \varepsilon.$$
\end{theorem}

This iterative regularization does not rely matrix inversion and therefore allows for a hardware efficient implementation. Note that the constraint $0 < \omega \leq \|A_S\|^{-2}_\text{op}$ can be satisfied by choosing
$\omega = \|A\|_F^{-2} \leq \|A_S\|_F^{-2} \leq \|A_S\|^{-2}_\text{op},$
which in turn is easy to compute. The Landweber 
regularized OMP (L-OMP) is summarized in Algorithm~\ref{alg:l-omp}. Note that the inner loop (Step~3b) is initialized with $x_S = 0$ in the first iteration of the outer loop. In the subsequent iterations, the inner loop takes the previous signal estimate $x_{S}$ as initialization for the Landweber iteration.


\begin{algorithm}
	\caption{L-OMP} \label{alg:l-omp}
	Input: sensing matrix $A \in \R^{m \times N}$, samples $y \in \R^m$, noise level $\varepsilon$, regularization parameter $\lambda \in \N^+$.
	
	Initialize: $r \leftarrow y$, $S \leftarrow \{\}$, $x \leftarrow 0 \in \R^N$.
	
	Until $\|r\|_2 \leq \varepsilon$:
	\begin{enumerate}
		\item Observe $A^*y$ and find the index $s$ of the largest element in magnitude, i.e.
		$$s \leftarrow \argmax_{j=1,...,N} |\langle A_j, r\rangle|.$$
		\item Add the element to the support, i.e. $S \leftarrow S \cup \{s\}$.
		\item Obtain new signal estimate by the iteration:
		\begin{enumerate}
			\item Initialize $0 < \omega \leq \|A_S\|_\op^{-2}$.
			\item For $\ell = 1,...,\lambda$:
			\begin{align*}
				x_S &\leftarrow x_S + \omega A_S^* (y - A_S x_S), \\
				x_{S^C} &\leftarrow 0.
			\end{align*}
		\end{enumerate}
		
		\item Update the residual by $r \leftarrow y - A_S x_S$.
	\end{enumerate}
\end{algorithm}

\textbf{Remark:} A similar approach can be found in \cite{F11}, where the same iterative scheme is used in the Fast Hard Thresholding Pursuit (FHTP). Similar ideas are discussed in \cite{NT09} for the CoSaMP. However, note that both algorithms rely on \textit{a priori} information on the sparsity level $k$ for the support augmentation step.

\section{Numerical results}

In this section, we evaluate the performance of the proposed T-OMP and L-OMP algorithms by a numerical experiment. We generate a random sensing matrix $A \in \R^{m \times N}$ where each entry is taken independently from a standard Gaussian distribution. Following \cite{LCHW17}, the columns of $A$ are then normalized. For the ambient dimension, we set $N=256$, and we consider numbers of measurement $m \in \{16, 32, 64\}$. The sparsity level is fixed at $k=8$, and the non-zeros entries of the signal vector $x \in \R^N$ are taken independently from a uniform distribution supported on $[-1, 1]$. The noise level is measured via the signal-to-noise ratio (SNR), given by
$$\text{SNR} = 10 \log_{10} \left( \frac{\|Ax\|_2^2}{\|v\|_2^2} \right).$$
Each component of the noise vector $v \in \R^m$ is sampled independently from a standard Gaussian distribution. The noise vector is scaled thereafter such that the desired SNR is attained. The reconstruction performance is measured by normalized root-mean-square error (NRMSE) defined by
$$\text{NRMSE} = \frac{1}{\sqrt{N}} \frac{\|x - \hat x\|_2}{\Delta},$$
where $x$ is the true signal, $\hat x$  is the reconstructed signal, and $\Delta$ is the spread between the largest and smallest entry of $x$. 

\begin{figure}[t]
	\centering
	\includegraphics[width=\figwidth]{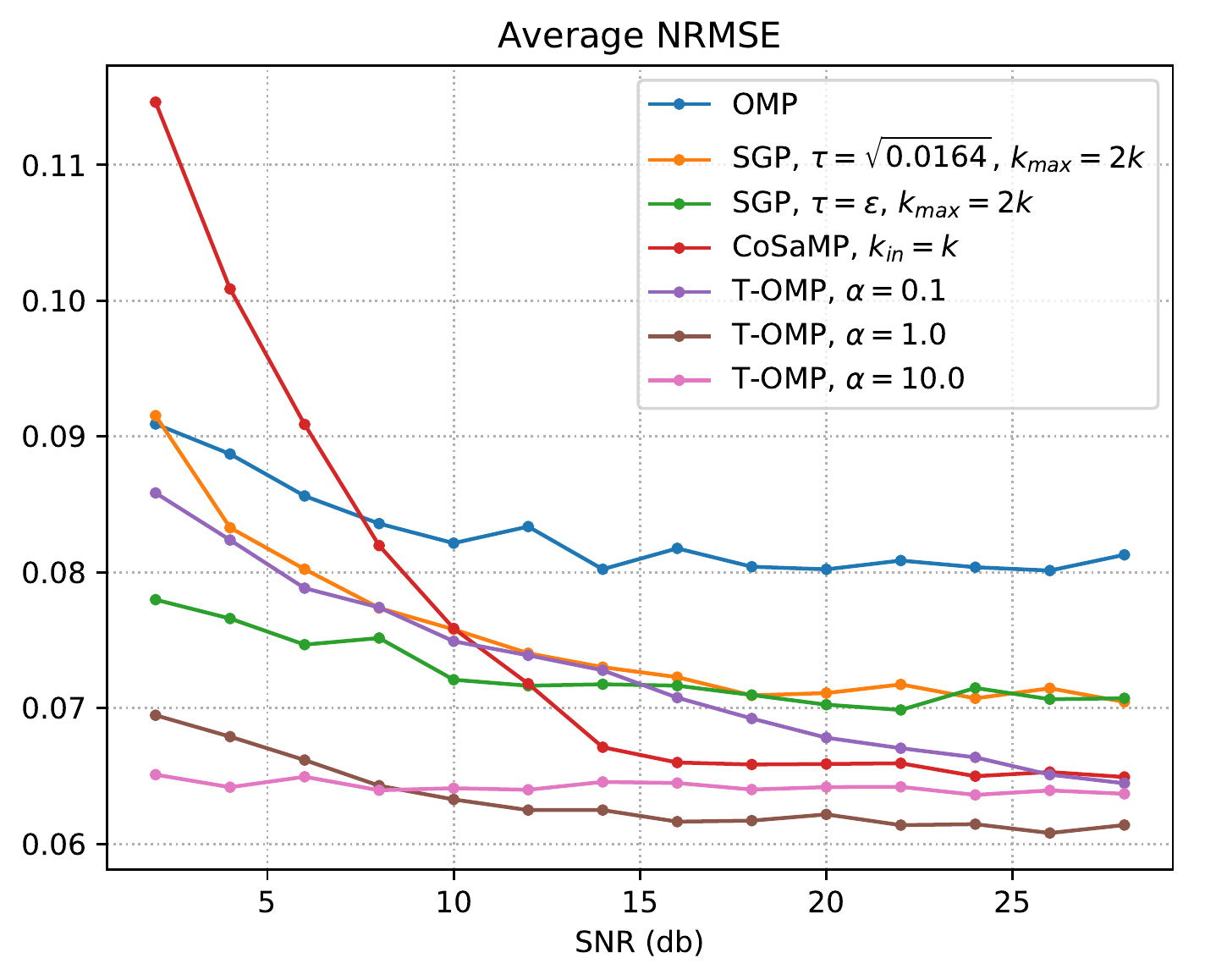}
	\includegraphics[width=\figwidth]{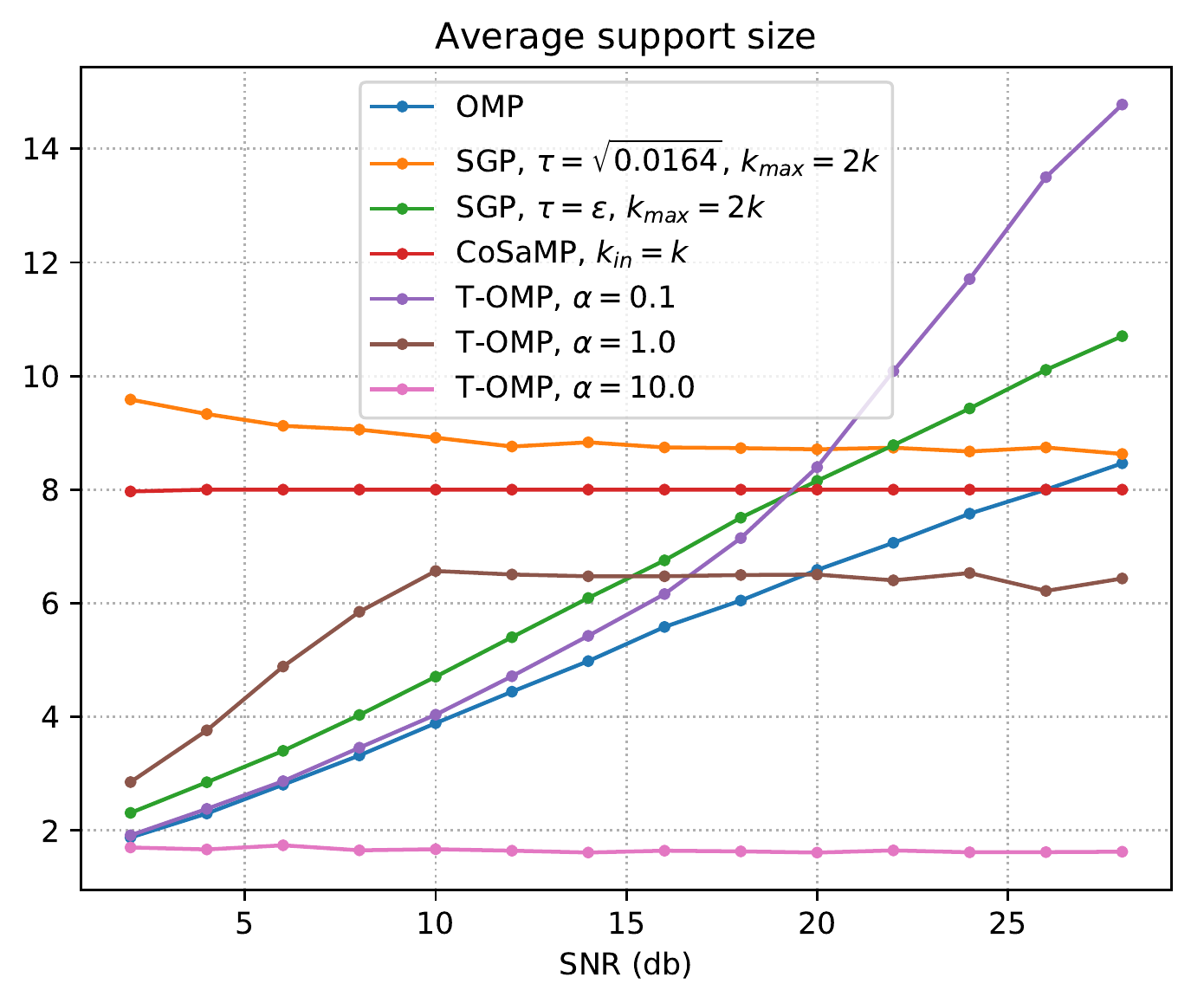}
	
	\caption{Reconstruction performance of the T-OMP algorithm with different regularization parameters $\alpha$ for different signal-to-noise ratios. The number of measurements is $\bm{m = 16}$. The curves for OMP, SGP, and CoSaMP serve as benchmark.} \label{fig:t-omp16}
\end{figure}

In order to make the presented algorithms' performance comparable, we employ the same stopping rule in all our experiments: The iteration is terminated as soon as $\|r\|_2 \leq \varepsilon$, where $r$ is the matching pursuit's residual, and $\varepsilon = \|v\|_2$ is the true energy of the generated noise. This stopping rule is somewhat of theoretical nature, since the true noise energy is unknown in applications. However, only a uniform halting criterion will allow for a fair competition between the algorithms assessed. Note that \citeauthor{LCHW17} propose a fixed halting criterion $\|r\|_2 \leq \sqrt{0.0164}$ which is independent from the magnitude of the noise. All implemented algorithms use the number of measurements $m$ as maximum number of iterations. The presented values are the average of 1,000 trials for each combination of method and noise level.

Figure~\ref{fig:t-omp16} displays the reconstruction performance of the proposed T-OMP algorithm for $m = 16$. The presented algorithms SGP, CoSaMP and T-OMP outperform OMP, with the exception of CoSaMP in the high noise regime. Looking at the SGP algorithm, the performance gap between parameter choice $\tau = \sqrt{0.0164}$ (as proposed in \cite{LCHW17}) and $\tau = \varepsilon$ highlights the importance of a uniform stopping criterion for a fair algorithm comparison. Interestingly, T-OMP with $\alpha = 1$ and $\alpha = 10$ reaches a stable estimated support size as the noise level decreases. The same holds true for SGP with a fixed, noise-independent $\tau$.

\begin{figure}
	\centering
	\includegraphics[width=\figwidth]{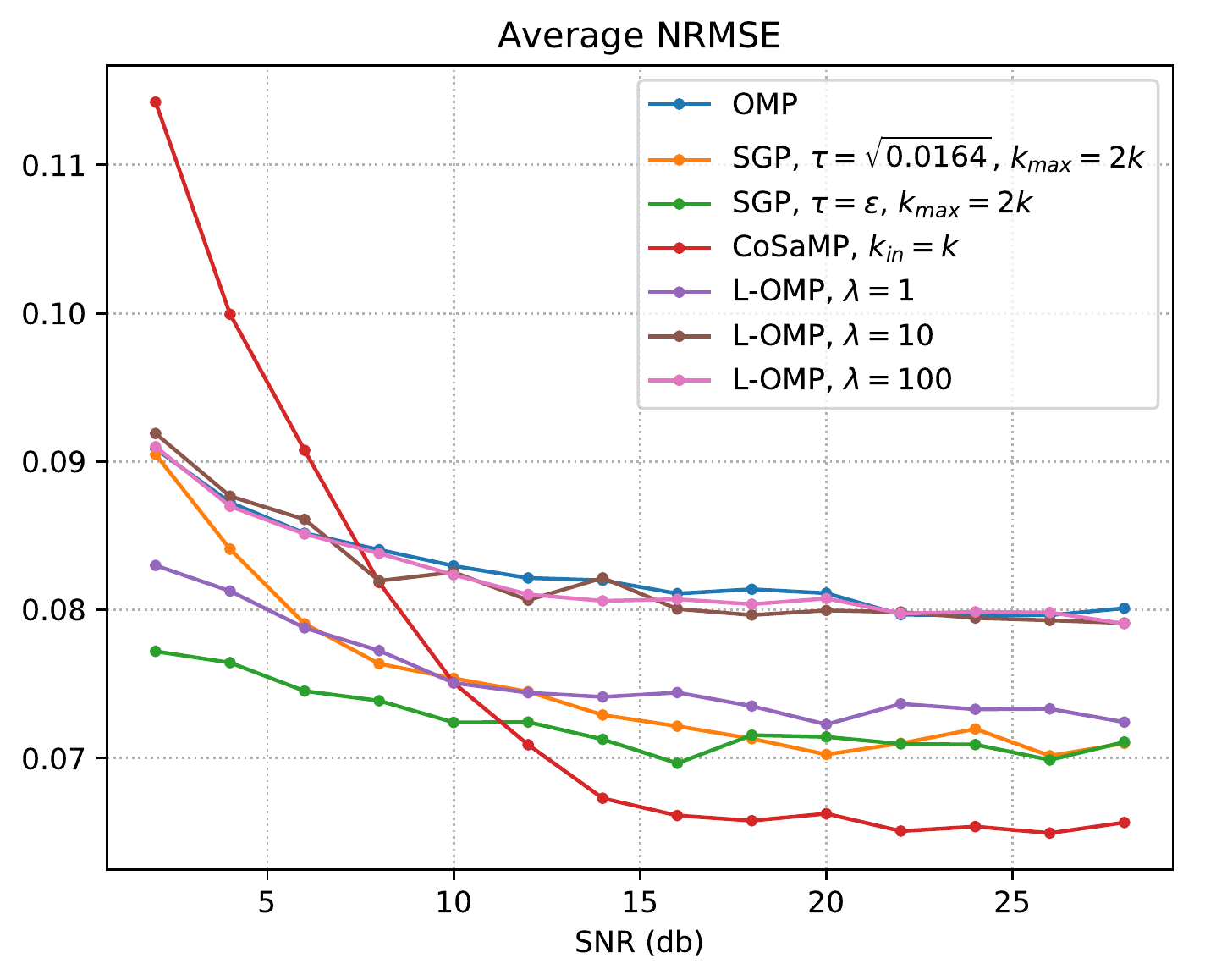}
	\includegraphics[width=\figwidth]{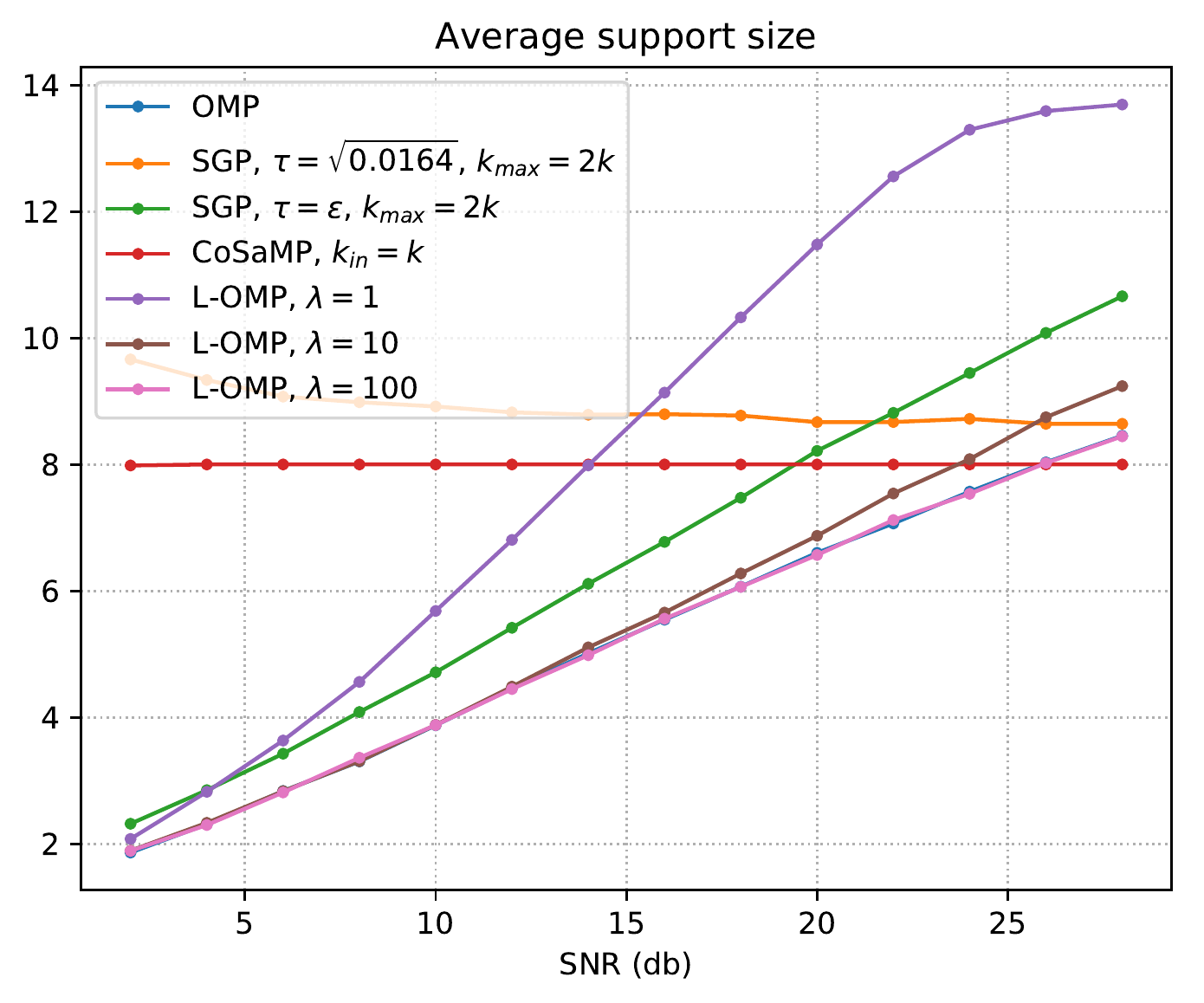}
	
	\caption{Reconstruction performance of the L-OMP algorithm with different regularization parameters $\lambda$ for different signal-to-noise ratios. The number of measurements is $\bm{m = 16}$. The curves for OMP, SGP, and CoSaMP serve as benchmark. Note that the curves for OMP (blue) and L-OMP with $\lambda = 100$ (pink) are often overlapping.} \label{fig:l-omp16}
\end{figure}

Figure~\ref{fig:l-omp16} displays the reconstruction performance of the proposed L-OMP algorithm in the same setting. For $\lambda = 100$, we observe that L-OMP produces results almost identical to the OMP algorithm as the pseudo-inverse is very closely approximated. Surprisingly, the Landweber method with only one iteration in the inner loop, i.e. $\lambda= 1$, finds a good reconstruction of the true signal in terms of NRMSE while it tends to misestimate the support size. For $\lambda = 10$, we observe a good approximation of the OMP solution which cannot outperform CoSaMP and SGP. However, note that SGP requires $m = 16$ iterations in its inner loop.

\begin{figure}
	\centering
	\includegraphics[width=\figwidth]{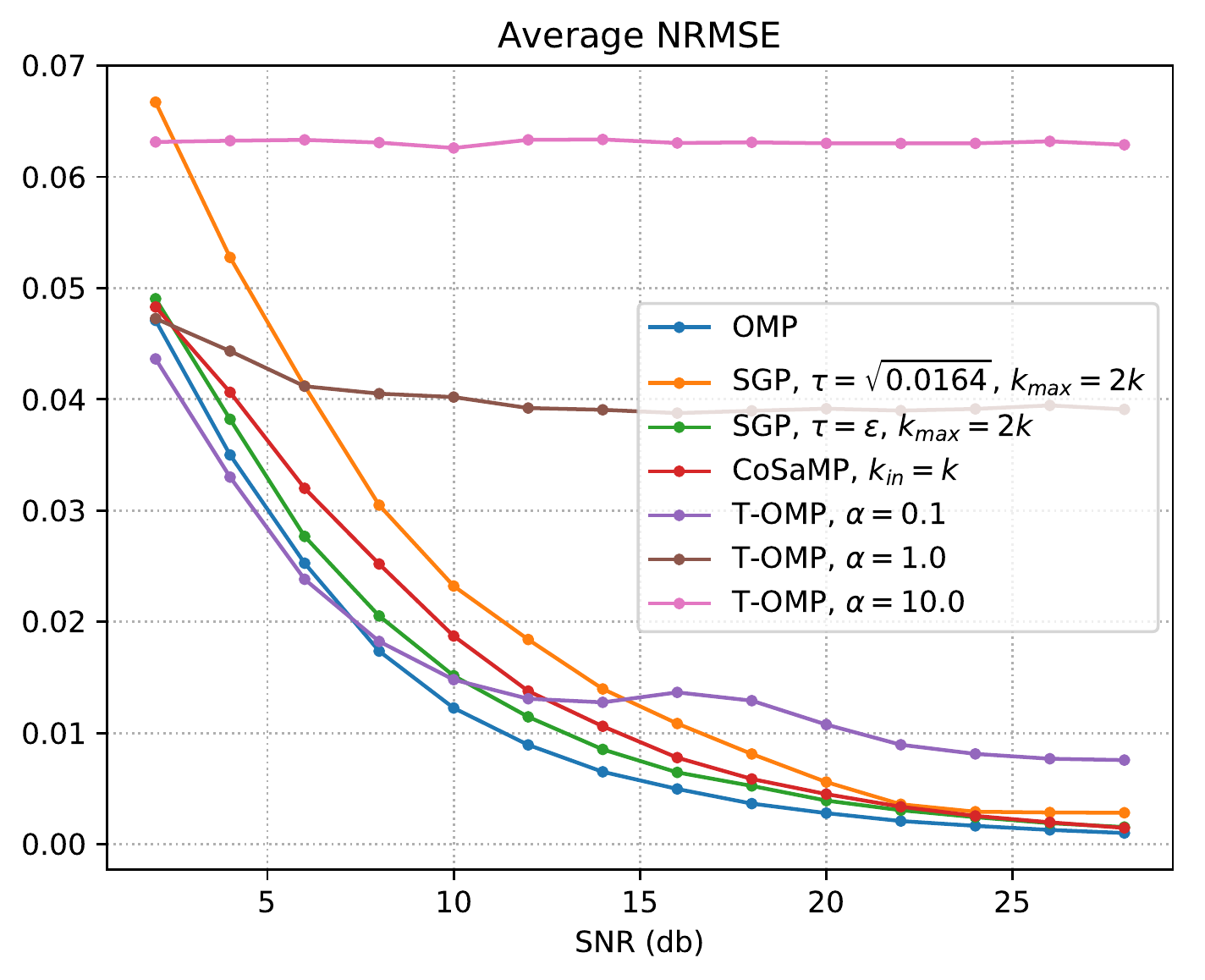}
	\includegraphics[width=\figwidth]{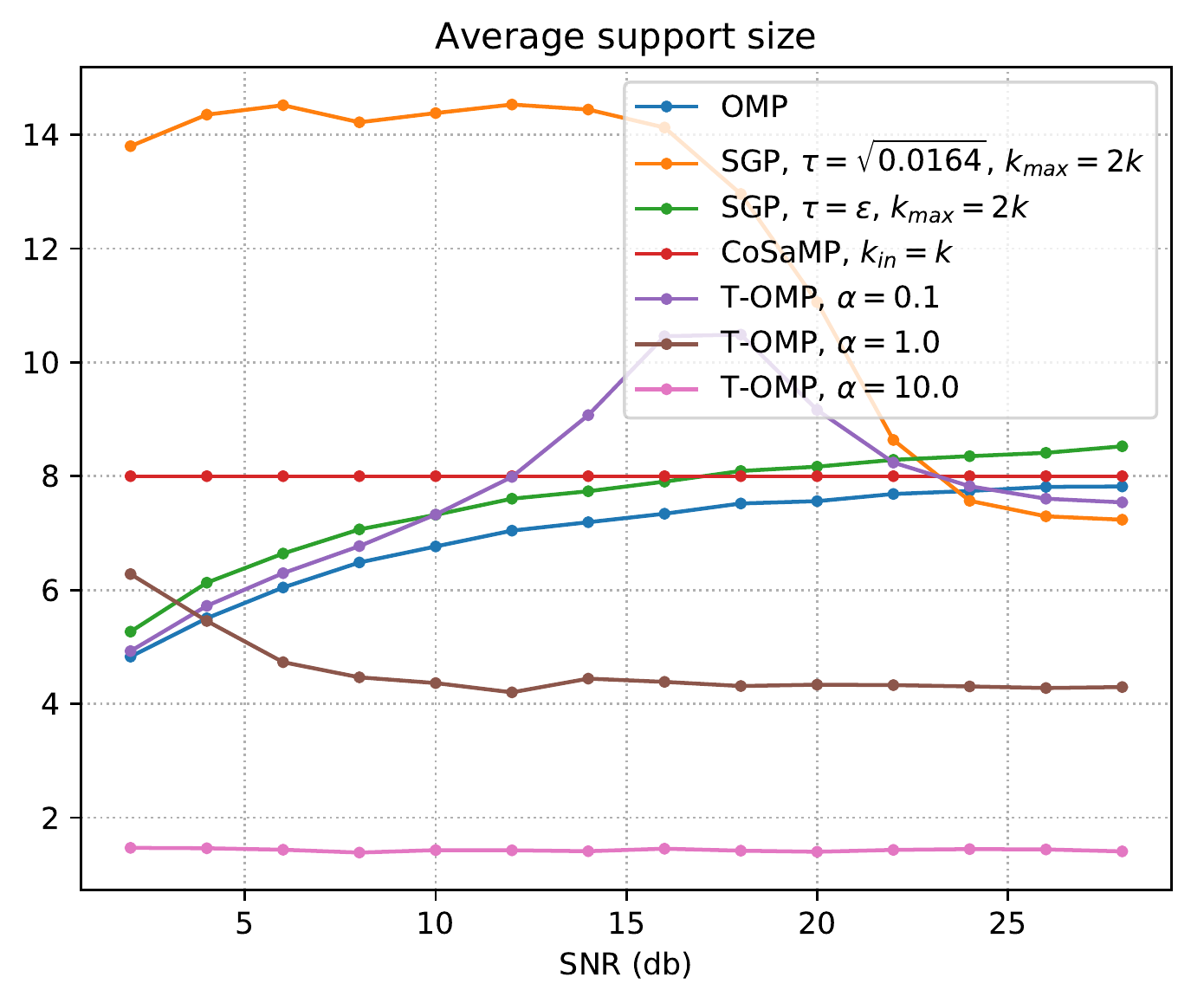}
	
	\caption{Reconstruction performance of the T-OMP algorithm with different regularization parameters for different signal-to-noise ratios. The number of measurements is $\bm{m = 64}$. The curves for OMP, SGP, and CoSaMP serve as benchmark.} \label{fig:t-omp64}
\end{figure}

\begin{figure}
	\centering
	\includegraphics[width=\figwidth]{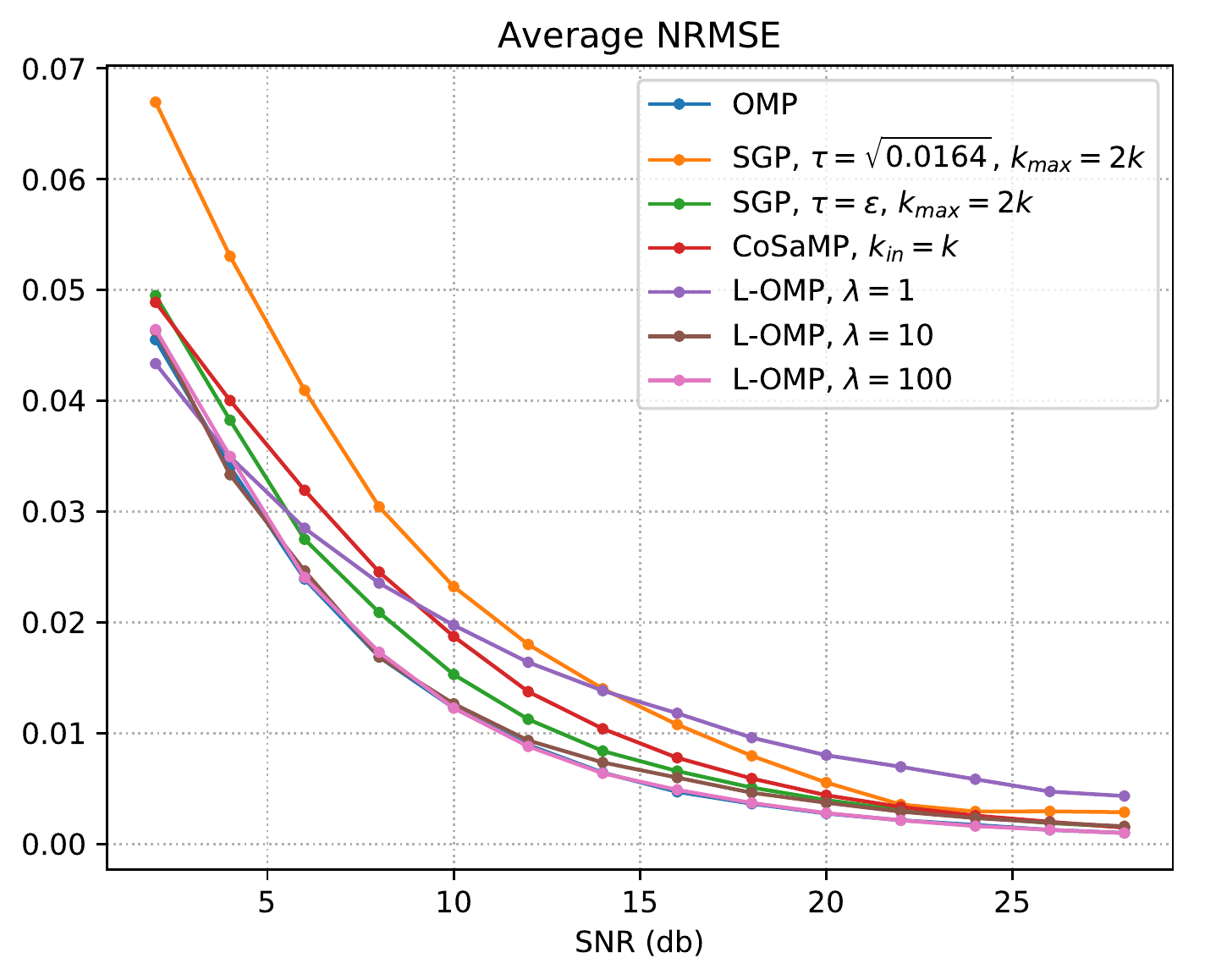}
	\includegraphics[width=\figwidth]{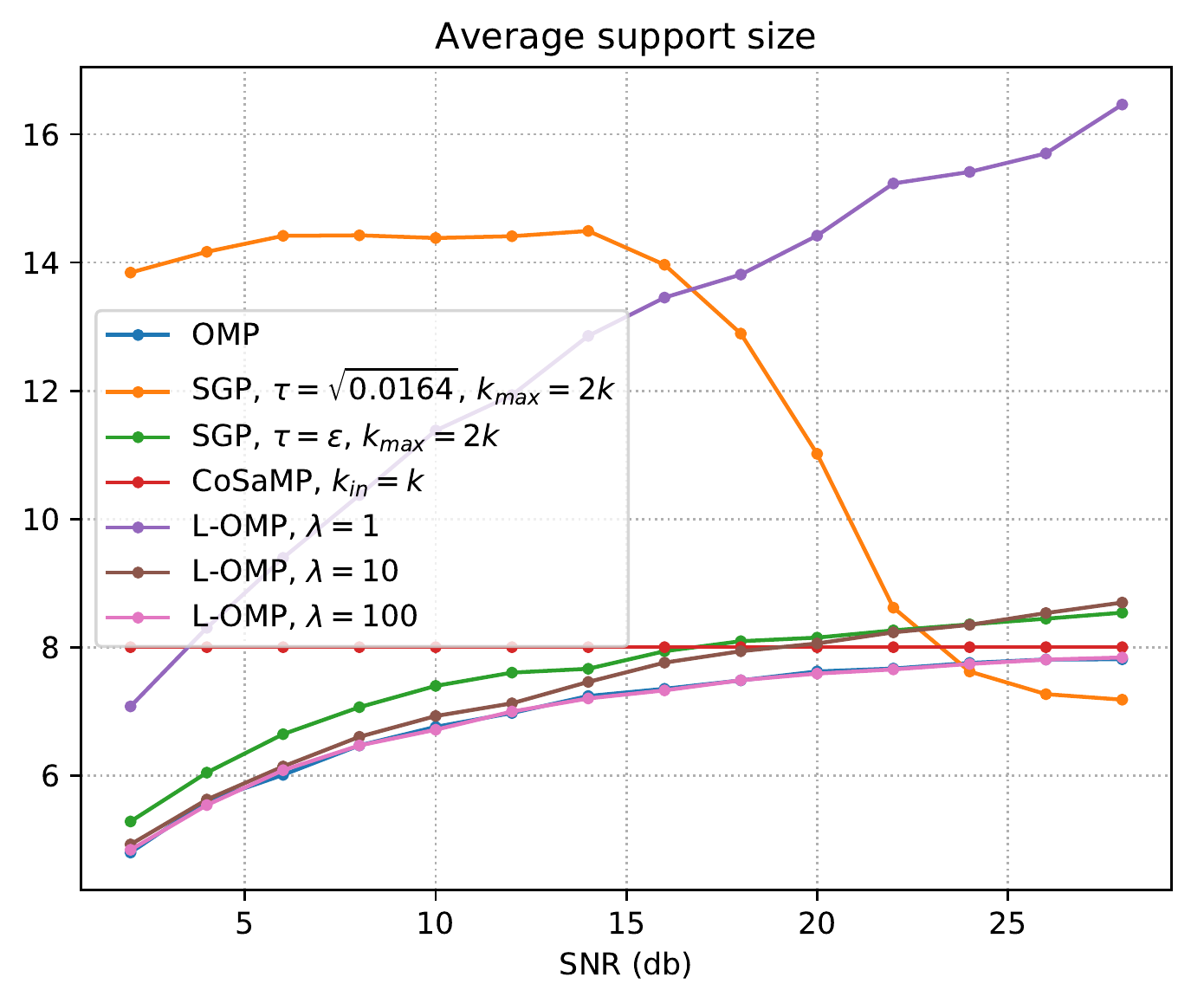}
	
	\caption{Reconstruction performance of the L-OMP algorithm with different regularization parameters for different signal-to-noise ratios. The number of measurements is $\bm{m = 64}$. The curves for OMP, SGP, and CoSaMP serve as benchmark. Note that the curves for OMP (blue) and L-OMP with $\lambda = 100$ (pink) are overlapping.} \label{fig:l-omp64}
\end{figure}

We now repeat the numerical experiments with a higher number of measurements, $m = 64$. In the light of (\ref{eqn:ric-m}), the random matrix $A_S$ is now more likely to have a good condition number. Furthermore, in the light of (\ref{eqn:omp-m}), the OMP algorithm has now a higher success probability. Indeed, Figures~\ref{fig:t-omp64}\nobreakdash--\ref{fig:l-omp64}\ confirm that the OMP algorithm cannot be outperformed on any noise level and regularization does not improve the reconstruction performance. Note that for an increasing SNR, the OMP now captures the correct support size. Looking at the iterative methods (Figure~\ref{fig:l-omp64}), the SGP tends to overestimate the support size. In this case, the Landweber method can be a good alternative as it approximates the OMP solution very well after only 10 iterations in the inner loop, while the SGP requires $m = 64$ iterations in its LMS step. Note that this setup reproduces the results of \cite{LCHW17}. Clearly, the proposed choice of $\tau = \sqrt{0.0164}$ in the SGP does not demonstrate the algorithm's full reconstruction capabilities.

\begin{figure}
	\centering
	\includegraphics[width=\figwidth]{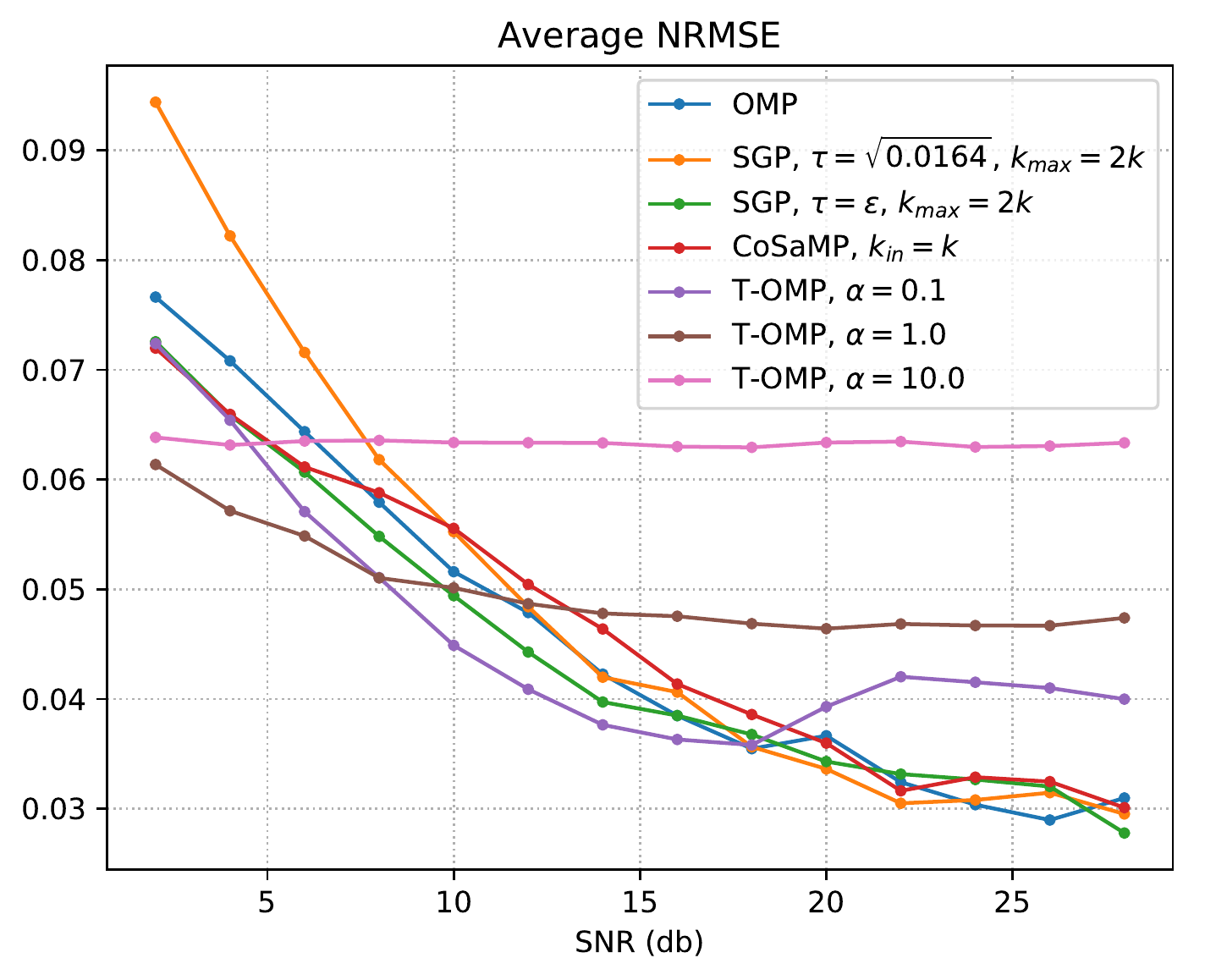}
	\includegraphics[width=\figwidth]{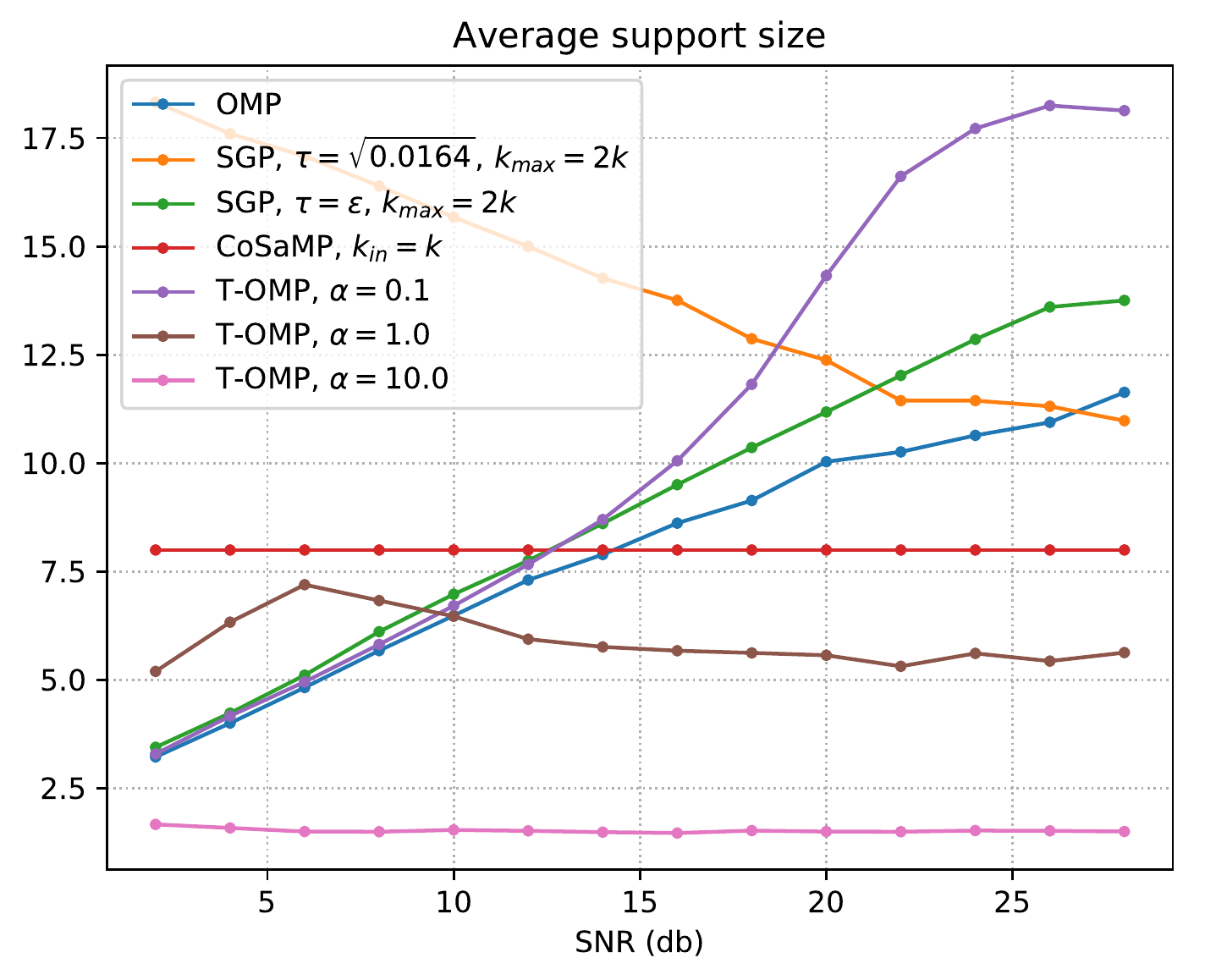}
	
	\caption{Reconstruction performance of the T-OMP algorithm with different regularization parameters for different signal-to-noise ratios. The number of measurements is $\bm{m = 32}$. The curves for OMP, SGP, and CoSaMP serve as benchmark.} \label{fig:t-omp32}
\end{figure}

\begin{figure}
	\centering
	\includegraphics[width=\figwidth]{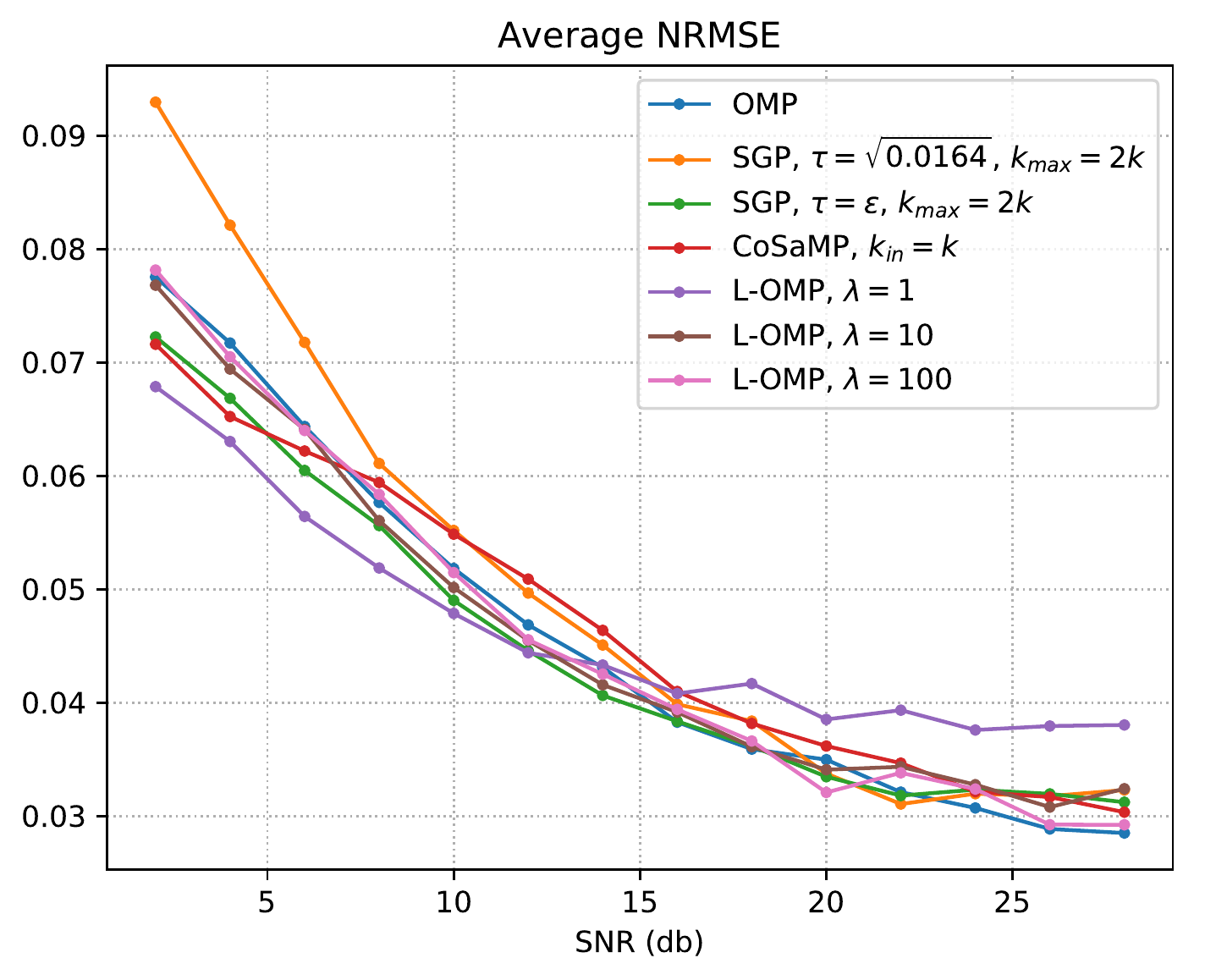}
	\includegraphics[width=\figwidth]{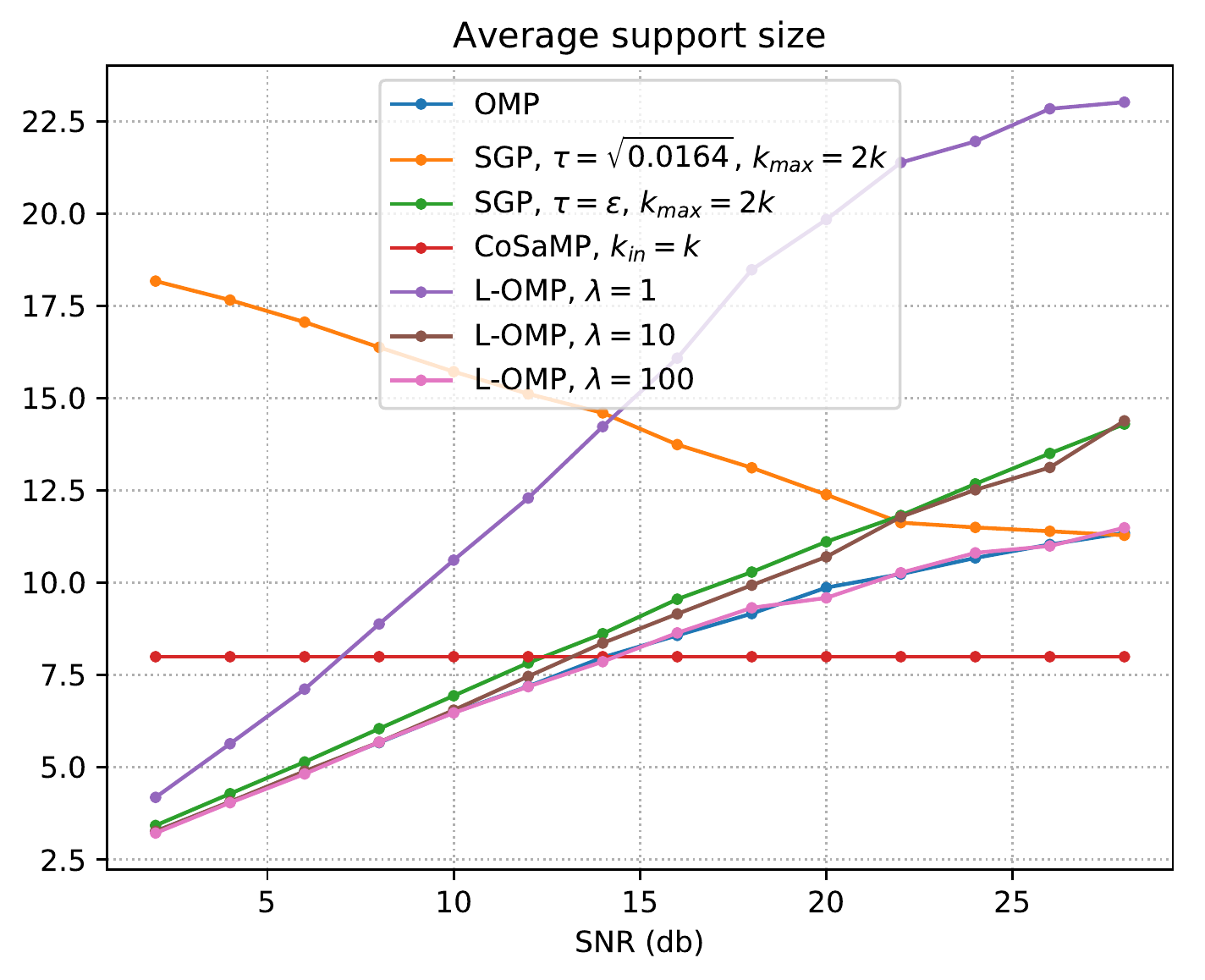}
	
	\caption{Reconstruction performance of the L-OMP algorithm with different regularization parameters for different signal-to-noise ratios. The number of measurements is $\bm{m = 32}$. The curves for OMP, SGP, and CoSaMP serve as benchmark. Note that the curves for OMP (blue) and L-OMP with $\lambda = 100$ (pink) are overlapping.} \label{fig:l-omp32}
\end{figure}

Finally, Figures~\ref{fig:t-omp32}\nobreakdash--\ref{fig:l-omp32} show the same simulation with $m = 32$ measurements. Both T-OMP and L-OMP outperform OMP in the high and medium noise regime. One can easily see that the choice $\alpha = 10$ leads to a poor overall performance due to a too strong regularization. T-OMP with $\alpha = 0.1$ has the best average success rates---at the price of an overestimated support. The L-OMP reaches the best overall performance with only one iteration ($\lambda = 1$), which demonstrates the advantages of regularized methods over the direct computation of the pseudoinverse.

\section{Conclusion}

In this work, we have derived two extensions of the OMP algorithm for compressed sensing based on Tikhonov regularization and Landweber iteration. A series of numerical experiments confirms the positive effect of regularization, especially in situations where the sampling matrix does not act like an almost-isometry on the set of sparse vectors. In particular, in situations where the number of measurements is comparably small, T-OMP and L-OMP outperform OMP and CoSaMP with a unified stopping criterion. Unlike CoSaMP, the proposed algorithms no not rely on \textit{a priori} information of the signal's sparsity. The L-OMP algorithm is an alternative iterative method for signal reconstruction that allows for a hardware-friendly implementation in the spirit of \cite{LCHW17}, as it renounces the computation of a pseudoinverse or the solution of a linear system, and shows good results after only few steps.

An important open question for applications are good parameter choice rules or heuristics for the regularization parameters in T-OMP and L-OMP, and halting criteria that lead to provable recovery guarantees. Parameter choice rules and halting criteria usually rely on \textit{a priori} information on the signal such as the sparsity or noise energy, such that the fine-tuning of T-OMP and L-OMP remains a highly application-specific problem.

The OMP algorithm and its robust modifications iterate over a scheme of \textit{support augmentation} and \textit{signal estimation}. With this modularity in mind, our regularization approach can be employed within the framework of other matching pursuits as well. For example, a study of CoSaMP or ROMP can be of future interest, where both the support augmentation (by the algorithms' corresponding techniques) and the signal estimation (by our proposed approach) are simultaneously regularized. By this fused approach, even higher reconstruction rates than proposed in this work can possibly be achieved.

\printbibliography

%
%
%
%
%
%
%
%
%
%
%

\end{document}